\documentclass[showpacs,aps]{revtex4}
\usepackage{amssymb}
\usepackage{amsmath}
\usepackage{graphicx}
\usepackage{lscape}
\usepackage{booktabs}
\begin{document}
\title{Study on space-time structure of Higgs boson decay using HBT correlation Method in e$^+$e$^-$ collision at $\sqrt{s}$~=~250 GeV}
\author{Hong-ge Xu$^{1}$, Di-kai Li\footnote{Corresponding Author:
lidikai@gmail.com,chengang1@cug.edu.cn}$^{1,2}$, Gang Chen$^1$,
Liang Li$^2$}
\address{
1 School of Mathematics and Physics, China University of Geoscience, Wuhan 430074, China\\
2 School of Physics, Shanghai jiaotong University, Shang Hai
430082,China}

\begin{abstract}
The space-time structure of the Higgs boson decay are carefully studied with the HBT
correlation method using e$^+$e$^-$ collision events produced through Monte Carlo generator PYTHIA 8.2 at $\sqrt{s}$~=~250~GeV. The Higgs boson jets (Higgs-jets) are identified by H-tag tracing. The measurement of the Higgs boson radius and decay lifetime are derived from HBT correlation of its decay final state pions inside Higgs-jets in the e$^+$e$^-$ collisions events with an upper bound of $R_H \le 1.03\pm 0.05$~fm and $\tau_H \le (1.29\pm0.15)\times 10^{-7}$~fs. This result is consistent with {\footnotesize CMS} data.

\end{abstract}
\pacs{25.75.-q, 24.85.+p, 24.10.Lx}

\maketitle

\section{Introduction}
The Standard Model (SM) of particle physics~\cite{1,2,3,4} has been tested by many experiments over the last four decades and has successfully described high energy particle interactions. However, the mechanism that breaks electro-weak symmetry in the SM had not been verified experimentally by then. In 1964, a new mechanism was proposed by several research groups to give the origin of the mass of elementary particles. This mechanism~\cite{5,6,7,8,9,10} implies the existence of a scalar particle, the SM Higgs boson. The search for this particle is a dominant part of the history of collider experiments in the last few decades. It has been through Large Electron Positron (LEP) Collider at CERN, the Tevatron at Fermilab, and the Large Hadron Collider (LHC). In the summer of 2012 the ATLAS Collaboration and the CMS Collaboration at CERN announced a new particle~\cite{11,12}, which is a Higgs-like boson.
The discovery of the Higgs boson has ushered in a new era of high energy physics. The Standard Model has been proved to be essentially correct, at least as a low-energy effective field theory, in its description of electroweak symmetry breaking as due to a light, weakly coupled scalar boson. However, the physics giving rise to the Higgs potential remains completely unclear. We expected that the three possible future colliders: the ILC~\cite{13}, FCC-ee (formerly known as TLEP)~\cite{14}, and CEPC (http://cepc.ihep.ac.cn) can offer clues to electroweak physics including the Higgs boson. In particular~\cite{15}. The CEPC e$^+$e$^-$ collider will bring a major leap in the precision measurement of the Higgs boson, and enable electroweak measurements with the Z and the W bosons. In the paper, we study the property of Higgs boson in e$^+$e$^-$ collisions at $\sqrt{s}$~=~250~GeV using the PYTHIA 8.2 generator and calculate the size of Higgs boson by HBT correlation.
HBT correlation as proposed by Hanbury Brown and Twiss~\cite{16,17}, the (angular) diameter of stars and radio sources in the universe was successfully determined by measuring the intensity correlations between separated telescopes. Likewise, in particle physics, one can in principle use Bose-Einstein correlations between identical particles to assess the spatial scale of the emitting source in high-energy collision. Bose-Einstein enhancement of identical-pion pairs at low relative momentum was first observed in $p\bar{p}$ collisions by Goldhaber, Goldhaber, Lee, and Pais 50 years ago~\cite{18}.

\section{Methodology}
\subsection{HBT correlation}
As we know, Hanbury Brown-Twiss analysis(HBT) has been successfully applied in e$^+$e$^-$~\cite{19}, hadron-hadron and lepton-hadron~\cite{20}, and heavy-ion~\cite{21}~collisions. The HBT correlation also called the Bose-Einstein correlation, which is the main method to measure the emission source size of final particles in high energy collision. Two-pion Hanbury-Brown-Twiss (HBT) interferometry is a powerful tool to study the space-time structure of particle-emitting sources produced in high energy collisions[22,23,24,25], most of the final state particles produced in e$^+$e$^-$ collisions are $\pi$ mesons. so we choose $\pi$ mesons ($\pi^+$, $\pi^-$ or $\pi^0$) as the identical particles to study.
The two-particle Bose-Einstein correlation function $C_{2}(k_{1},k_{2})$ is defined as the ratio of the two-particle momentum distribution correlation function $P(k_1,k_2)$ to the product of the  single-particle momentum distribution $P(k_1)P(k_2)$. For an expanding source, $P(k)$ and $P(k_1,k_2)$ can be expressed as
\begin{equation}
P(k) = \int{dx \rho(x) A^{2}(k,x)},
\end{equation}
 \begin{center}
\begin{eqnarray}
&P(k_1,k_2)= \frac{1}{2!}{|\Psi(k_{1},k_{2};x_{1}',x_{2}')|}^{2}\ \ \ \ \ \ \ \ \ \ \nonumber\\
  &= P(k_1)P(k_2)+{|\int{dx e^{i(k_{1}-k_{2})x}\rho(x;k_{1},k_{2})}|}^{2},
\end{eqnarray}
\end{center}
where $A(k,x)$ is the amplitude for producing a pion with momentum k at x, $\rho(x)$ is the pion-source density, $k_{1}$, $k_{2}$, $x_{1}'$, $x_{2}'$ is the momentum and detecting points of two identical pions.
If we introduce effective density function $\rho_{eff}$, the  $P(k_1,k_2)$ will be expressed as
\begin{equation}
P(k_1,k_2) = P(k_1)P(k_2)+{|\int{dx e^{i(k_{1}-k_{2})x}\rho_{eff}(x;k_{1},k_{2})}|}^{2},
\end{equation}
or
\begin{equation}
P(k_1,k_2) = P(k_1)P(k_2)(1+{|\widetilde{\rho}_{eff}(q;k_{1}k_{2})|}^{2}),
\end{equation}
Here,
\begin{equation}
\rho_{eff} = \frac{\rho(x)(k_{1},x)(k_{2},x)}{\sqrt{(P(k_1)P(k_2))}}
\end{equation}
\begin{equation}
\widetilde{\rho}_{eff} = \int{dx e^{iq.x}\rho_{eff}(x;k_{1}k_{2})}
\end{equation}
where $q = k_{1}-k_{2}$ is the four-dimensional momentum difference, the $\widetilde{\rho}_{eff}$ is the Fourier transform of $\rho_{eff}$. So the correlation function  $C_{2}(k_{1},k_{2})$ can be written as
\begin{equation}
C_2(k_1,k_2) = \frac{P(k_1,k_2)}{P(k_1)p(k_2)}\\=1+{|\widetilde{\rho}_{eff}(q;k_{1}k_{2})|}^{2},
\end{equation}
 If the effective density function of the source is parameterized to Gaussian form, we have
\begin{equation}
C_2(q)=C_2(q;k_1,k_2)\\=1+{\lambda}e^{-R_{x}^{2}q_{x}^{2}-R_{y}^{2}q_{y}^{2}-R_{z}^{2}q_{z}^{2}-\sigma_{t}^{2}q_{t}^{2}}
\end{equation}
where $q = k_{1}-k_{2}$ is the four-dimensional momentum difference. If only assuming the distribution of the source is isotropic, i.e. the distribution function of the source is taken as spherically symmetric, the correlation function can be simplified as~\cite{26,27}~
\begin{equation}
C_2(q)=C_2(Q,q_0)\\=1+{\lambda}e^{-Q^{2}R^{2}-q_0^{2}\tau^{2}}
\end{equation}
Here $\lambda$ is the incoherence parameter, in the range $0\leq\lambda\leq1$, $\lambda$ denotes the correlation strength; R is the source radius, and $\vec Q = {\vec p_1}-{\vec p_2}$ is the three-dimensional momentum difference, $q_0 = |E_1-E_2|$ is the energy difference, $\tau$ is the decay-life. In this paper, we study the average radius R and decay-life of the Higgs bosons through the correlation function of the pion source production from Higgs decay in e$^+$e$^-$ collisions, which is taken as spherically symmetric. Then, the information about the average size and the decay-life of the emitting source for the final state $\pi$ mesons can be obtained. The two-particle correlation function with statistical method is defined as the ratio
\begin{equation}
C_2(k_1,k_2)=C_2(Q,q_0)=\frac{A(Q,q_0)}{B(Q,q_0)},
\end{equation}
where the two particles with momenta difference $Q$ and energy difference $q_0$ are from one same event, $A(Q,q_0)~=~ {\Delta{N}}/{\Delta{q_{corre}}}$ is the four-dimensional distribution function of the identical particles with HBT correlations, and  $B(q)~=~ {\Delta{N}}/{\Delta{q}}$ is the four-dimensional distribution function of the identical particles without HBT correlations. The momentum difference of the $\pi$ meson pairs are calculated. The correlation among identical particles with large momentum difference is quite weak, the distribution here with the HBT correlation should be the same as the distribution without the HBT correlation.
\subsection{Particle trace}
The particular importance of HBT correlation method is to identify identical particles from the same emitting source. Various methods have been proposed, among which the Higgs-tag that originated form b-tag method~\cite{28}~is a good choice. We use Monte Carlo generator PHYTHIA 8.2 to simulate e$^+$e$^-$ collision events both with and without Bose-Einstein effect, and then select suitable events for study. We trace the Higgs decay process and all daughters of Higgs boson and select all pions of the daughters as the identical particle from the same emitting source. Then, identical $\pi$ mesons are selected from the final state particles to make pion pairs after any two $\pi$ mesons are grouped with each other.
\subsection{Monte Carlo simulation}
We produce e$^+$e$^-$ collision events at $\sqrt{s}$~=~250~GeV using Monte Carlo generator PYTHIA 8.2. All production and decay channels of the SM Higgs boson are taken into account, setting the parameters of branching ratios as described in SM. The Higgs width value is $0.00403$ which is based on LHC data~\cite{29}. The other parameters are fixed on the default values given in PYTHIA. Specifically, for the parameters of hadrons production we set Parton Level~:~MPI = off, PartonLevel~:~ISR = off and PartonLevel~:~FSR = off.

\section{Result}
\subsection{Higgs Jet property}
Based on the MC simulation sample of one million e$^+$e$^-$ collision events at $\sqrt{s}$~=~250~GeV~using PYTHIA 8.219, we analyze among various SM Higgs boson production channels. We select different decay processes from total 1000,000 e$^+$e$^-$ collision events then analyze the property of jets in each Higgs boson decay process. All observable final-state particles, i.e. excluding neutrinos and other particles with no strong or electromagnetic interactions, are considered for analysis from the Monte Carlo events. We use cluster algorithms to predict the property of jets from Higgs boson in c.m. frame of e$^+$e$^-$ collisions events. The usage of cluster algorithms for e$^+$e$^-$ applications started in the late 1970¡¯s. A number of different approaches were proposed, We choose distance measure i.e. Lund model~\cite{30}. If the distances between the two nearest clusters is smaller than some cut-off value, the two clusters are joined into one.In this approach, each single particle belongs to exactly one cluster. Also note that the resulting jet picture explicitly depends on the cut-off value used. Based on the distance calculation, we set the jets cutting off parameter rapidity and transverse momentum is 0.01 and 10.0 respectively. Our leading jet is the one with the highest transverse momentum in each event. Our second leading jet is the one with the second transverse momentum in each event.The Dijet mass is:
\begin{equation*}
M=\sqrt{{M_1^2}+{M_2^2}+2({E_1}{E_2}-{\vec p_1}{\vec p_2})}
\end{equation*}

where $m_1$ and $m_2$ is leading jet's and second leading second jet's mass, $E_1$ and $p_1$ is leading jet's energy and momentum, $E_2$ and $p_2$ is second leading jet's energy and momentum. We calculate the Dijet mass of different Higgs decay channel.

\begin{figure}
\includegraphics[width=0.4\textwidth]{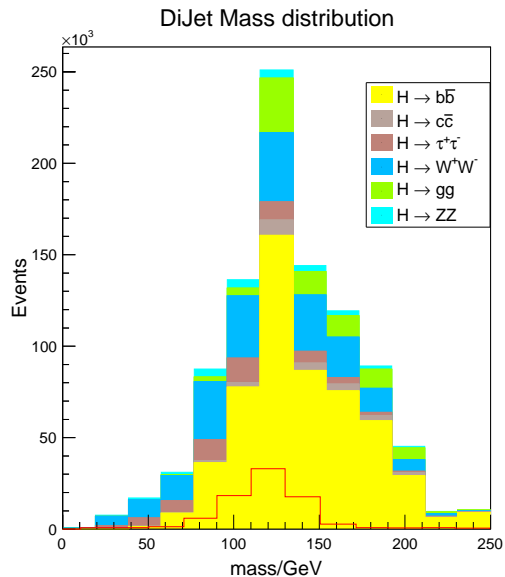}
\caption{Invariant mass of the leading and second-leading jets in e$^+$e$^-$ collision events. The different color represent different Higgs boson decay processes. The red line present experimental data multiplied by 200 given by D0 Collaboration~\cite{31}.}
\label{tu1}
\end{figure}

Figure~\ref{tu1} shows the distribution of the invariant mass of Dijet from different Higgs boson decay processes, including $H\rightarrow{b\bar{b}}$, $H\rightarrow{c\bar{c}}$, $H\rightarrow{\tau^+\tau^-}$, $H\rightarrow{W^+W^-}$, $H\rightarrow{gg}$, $H\rightarrow{ZZ}$. We extract the signal data published by D0 Collaboration~\cite{31}. There is an obvious peak around 125 GeV, which is the SM Higgs boson mass we choose. The distribution of Dijet mass, from our Monte Carlo e$^+$e$^-$ events sample, has a peak at 125 GeV, that is consistent with D0 experiment. Thus, we can convince our further study of Higgs boson is based on a viable PYTHIA simulation sample.

\subsection{The space-time structure of Higgs decay}

We use Monte Carlo generator PYTHIA 8.219 to produce  statisfied Higgs decay processes at $\sqrt{s}~=~250$~GeV. The event samples are constructed according to the different Higgs boson decay channel, including $H\rightarrow{b\bar{b}}$, $H\rightarrow{c\bar{c}}$, $H\rightarrow{g g}$, $H\rightarrow{all}$~(all may decay processes, not only above three processes), and to the cases with and without Bose-Einstein correlation effect by setting the parameter BoseEinstein:Pion = on/off, respectively. The subsample of each Higgs boson decay channel have 2000,000 e$^+$e$^-$ collision events and equally divided the two million events of each decay process in ten groups. We select and constitute Higgs-jets using an Higgs-tag method from an e$^+$e$^-$ collision events including Higgs boson decay and identical pions ($\pi^{+}$, $\pi^{-}$ and $\pi^{0}$) in Higgs-jets.

\begin{figure}[!htb]
\includegraphics[width=0.4\textwidth]{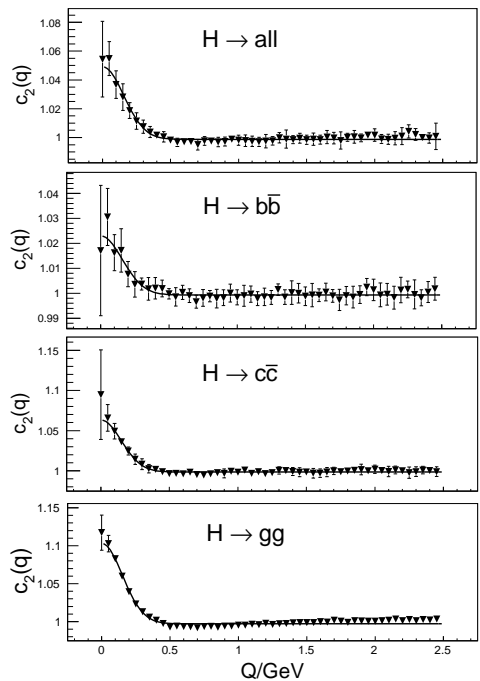}
\caption{The distribution of correlation functions for pion mesons inside Higgs-jets
with different decay channel from e$^+$e$^-$ collision events at $\sqrt{s}=250$~GeV,
including  $H\rightarrow{all}$, $H\rightarrow{b\bar{b}}$, $H\rightarrow{c\bar{c}}$, $H\rightarrow{gg}$.
The curves represent the results of fitting data by Eq.9.}
\label{tu2}
\end{figure}

Then according to Eq.(10), we calculate the correlation functions of identical pions ($\pi^{+}$, $\pi^{-}$ and $\pi^{0}$) from Higgs-jets of different decay channel and their standard deviation, as shown in Fig.2. Here, we choose the momentum and energy interval region is $Q=0\thicksim2.5$~GeV, $q_0 \le 15$MeV/c and equally divide the region into 50 bins, since the correlation among identical particles with large momentum difference is quite weak. The average radius size $R_{Hj}$ and decay lifetime $\tau_{Hj}$ of the Higgs boson (emitting source) can be obtained by fitting the correlation functions, the radius and decay lifetime values are filled in the table 1. The fit curves as Eq.(9) are shown in Figure~\ref{tu2}.

Figure~\ref{tu2} shows that the distribution of pion mesons correlation functions from different decay channel inside Higgs-jets of e$^+$e$^-$ collision events are similar. The average radius size $R_{Hj}$ and decay lifetime $\tau_{Hj}$ of the emitting source of pions from different Higgs boson decay mode can be obtained by fitting the correlation functions distributions in Figs.2 through formula(9). Here the radius value $R_{Hj}$ representing the size of the emission source of jets decaying from Higgs boson, including the size of the Higgs and the scale of the parton cascade process from Higgs decay before hadronization. So the radius of Higgs boson $R_H \le R_{Hj}$. Similarly, the decay lifetime here $\tau_{Hj}$ also contains in addition to Higgs after the disintegration of the secondary decay time, so real decay lifetime of Higgs $\tau_{H} \le \tau_{Hj}$.

\begin{figure}[!htb]
\includegraphics[width=0.5\textwidth]{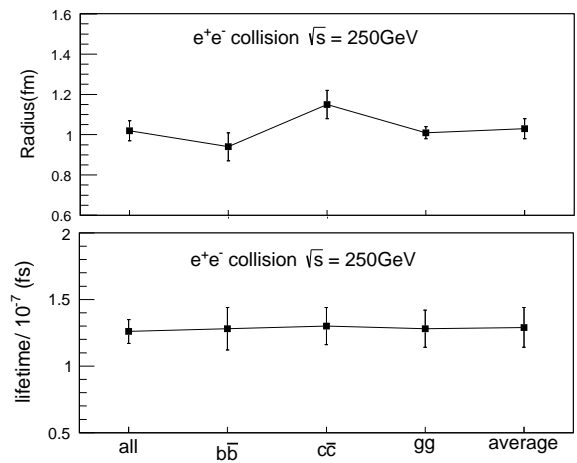}
\caption{ The decay radius values $R_{Hj}$, decay lifetime values $\tau_{Hj}$ and their average values of Higgs-jets source from different decay channel in e$^+$e$^-$ collision events at $\sqrt{s}=250$~GeV, measured by HBT correlations method using pions inside Higgs-jets.}
\label{tu3}
\end{figure}

For the convenience of comparison, the radius values $R_{Hj}$ and decay lifetimes $\tau_Hj$ of the measurement difference decay channel and their average values are filled in the Figure~\ref{tu3}, respectively. Figure~\ref{tu3} shows that space-time structure character i.e. the decay radius values $R_{Hj}$ and the lifetime value $\tau_{Hj}$ of Higgs-jets measured from different decay channels are same within the error range. The average radius of Higgs-jets source is $1.03\pm0.05$ fm and the average decay lifetime is $(1.29\pm0.15)\times 10^{-7}$~fs.
This is just fall within the scope of the {\footnotesize CMS} experimental results~\cite{32}.

\begin{table*}[tbp]
\caption{The fit parameter values of Higgs boson decay measured by HBT correlation method using pions from three main decay processes and the all decay process.}
\begin{tabular}{ccccccccc}
\hline  \hline
\quad Decay processes \quad& \quad$H\rightarrow{all}$\quad\quad&\quad$H\rightarrow{b\bar{b}}$  \quad\quad&\quad$H\rightarrow{c\bar{c}}$  \quad\quad&\quad$H\rightarrow{gg}$ \quad\quad&average \\
\hline
Braching Ratio&1&  0.578&  0.0268&  0.0856\\
Radius $R_{Hj}$(fm)&1.02$\pm$0.03&  0.94$\pm$0.07&  1.15$\pm$0.07&  1.01$\pm$0.03&  1.03$\pm$0.05\\
lifetime $\tau_{Hj}~(\times 10^{-7}$fs)&1.26$\pm$0.09 &1.28$\pm$0.16  &1.30$\pm$0.14&1.28$\pm$0.14&1.29$\pm$0.15\\
\hline  \hline
\end{tabular} \label{paci1}
\end{table*}
\section{Summary and discussion}
We use Monte Carlo simulation generator PYTHIA 8.219 to produce the data
of e$^+$e$^-$ collision events at $\sqrt{s}$~=~250~GeV, according to the different Higgs boson decay channel, including $H\rightarrow{all}$, $H\rightarrow{b\bar{b}}$, $H\rightarrow{c\bar{c}}$, $H\rightarrow{gg}$, respectively. The Higgs-jets are selected using the H-tag method. The space-time structure character of decay and evolution of Higgs boson are studied in detail by the HBT correlation method.

Firstly, we calculated the invariant mass of Dijet from Higgs boson in  e$^+$e$^-$ collisions and the results are consistent with the experimental data. Then the values of the average radii and decay lifetime of Higgs-jets source are measured by the HBT correlation method. We found that the average radii or decay lifetime of Higgs-jets source are the same within the error range at the different decay channel, including $H\rightarrow{all}$, $H\rightarrow{b\bar{b}}$, $H\rightarrow{c\bar{c}}$, $H\rightarrow{gg}$, respectively. So the mean value of radius and lifetime of Higgs-jets source from the different decay channel are obtained in e$^+$e$^-$ collisions at $\sqrt{s}=250$~GeV, i.e $R_{Hj} = 1.03\pm0.05$~fm and $\tau_{Hj} = (1.29\pm0.15)\times 10^{-7}$~fs. In addition, it is pointed out that the source
radii and/or decay lifetime of Higgs-jets measured at different decay channels are the same within the error range, which would mean that the average source radii
of Higgs-jets just reflect some intrinsic properties of Higgs.

It is worth noting that here the radius $R_{Hj}$ and/or decay lifetime $\tau_{Hj}$ of Higgs-jets source obtained do not equal to the radius $R_{H}$ and/or decay lifetime $\tau_{H}$ of Higgs.
Because from Higgs decay to form Higgs-jet should pass Higgs decay and hadronic secondary decay process, the radius $R_{Hj}$ and/or decay lifetime $\tau_{Hj}$ of Higgs-jets source are greater then the radius $R_{Hj}$ and/or decay lifetime $\tau_{Hj}$ of Higgs. So we present an upper bound of Higgs boson radius $R_H \le 1.03\pm 0.05$~fm and decay lifetime $\tau_H \le (1.29\pm0.15)\times 10^{-7}$~fs using from HBT correlation of its decay final state pions inside Higgs-jets in the e$^+$e$^-$ collisions events.
This result is consistent with {\footnotesize CMS} data~\cite{32}.
We also expect that this results will be tested in CEPC experiments in the future.

\section {ACKNOWLEDGMENT}
Finally, we acknowledge the financial support from NSFC(11475149).

\end{document}